\documentclass[preprint,prb,showpacs,amsmath,amssymb,superscriptaddress]{revtex4}
\usepackage{amsfonts,amsmath,mathrsfs,epsfig}
\begin{document}
%\preprint{PREPRINT}

\newcommand{\ud}{\mathrm{d}}

\title{Theory of domain formation in inhomogeneous ferromagnetic dipolar condensates }

\author{Jay D. Sau}
\email{jaydsau@umd.edu}
\affiliation{Department of Physics, University of Maryland,
College Park, Maryland 20742, USA}
\affiliation{Materials  Sciences Division,
Lawrence Berkeley National Laboratory, Berkeley, California 94720, USA}
\author{S. R. Leslie}
\affiliation{Department of Physics, University of California,
Berkeley, California 94720, USA}
\author{D. M. Stamper-Kurn}
\author{Marvin L. Cohen}
\affiliation{Department of Physics, University of California,
Berkeley, California 94720, USA}
\affiliation{Materials  Sciences Division,
Lawrence Berkeley National Laboratory, Berkeley, California 94720, USA}

\date{\today}

\begin{abstract}
Recent experimental studies of  $^{87}$Rb spinor Bose Einstein condensates  have shown the
existence of a magnetic field driven quantum phase transition accompanied by  structure formation on the
ferromagnetic side of the transition. In this theoretical study we
examine the dynamics of the unstable modes following the transition taking into account the
effects of the trap, non-linearities, finite temperature and dipole-dipole interactions. Starting from an initial state
which includes quantum fluctuations, we attempt to make quantitative comparisons with recent
experimental data on this system and estimate the contribution of quantum
zero-point fluctuations to the domain formation. Finally, using the strong anisotropy of the
trap, we propose ways to  observe directly the effects of dipole-dipole interactions on the spinor condensate
dynamics.
\end{abstract}

\maketitle
\section{Introduction}

The physics of phase transitions between ordered and disordered phases
 has been studied extensively
in the past and has yielded  answers to many fundamental questions about the effects of interactions and of quantum and
thermal fluctuations
on the equilibrium properties  in various phases.
 However, because of  the short time-scales and high levels
of noise and impurities involved in traditional condensed matter systems, it has been difficult
to study the \textit{dynamics} of such transitions in detail. Ultra-cold atoms, because of  their low
densities and temperatures, provide an opportunity to study the non-equilibrium dynamics
around phase transitions in a spatially and temporally resolved fashion. Bose Einstein condensates (BECs) of atoms with a
spin degree of freedom, i.e.\ spinor BECs, are an example of such a system where it is possible
to observe  non-equilibrium spin dynamics and how they are affected by quantum noise and the proximity to phase
transitions\cite{dmsk_nature}.

Recent experiments \cite{dmsk_nature,dmsk_prl} have reported the formation of magnetic structures
in  ultra-cold spin-1 $^{87}$Rb gases. Spin-1 atoms support two characteristic families of quantum states: polar states
exemplified by the $m_z=0$ hyperfine state and magnetic states exemplified by the $m_z=\pm 1$ hyperfine state\cite{ho,machida,barnett}; here $m_z$ denotes
the eigenvalue for
the dimensionless spin projection along the $\hat{z}$ axis.
As discussed further below, spin-dependent contact interactions naturally favor the ferromagnetic state in $^{87}$Rb spinor
condensates. The Rb atom may also be subjected to an extrinsic quadratic Zeeman shift, which lowers the
energy of the $m_z=0$ state by an amount $q$ with respect to the average of the energies of the $m_z=\pm 1$
states. In the experiments, a non-magnetic $m_z=0$ condensate is prepared at a large value
of $q$, where its internal state composition is stable. Following this, $q$ is rapidly
quenched to a regime where the initially prepared $m_z=0$ becomes unstable. As population flows to the
initially unoccupied $m_z=\pm 1$ states, the condensed gas is   observed to
  break translational and rotational symmetry spontaneously and form ferromagnetic domains of transversely
magnetized atoms. In the experiment, the system also evolves under  a significant linear Zeeman splitting of the three
Zeeman sublevels $m_z=0,\pm 1$. However, as we shall see later, since most of the terms of the
Hamiltonian are invariant under global unitary spin rotations, the linear Zeeman shifts can be eliminated from the theoretical treatment by a global unitary
 transformation of the spin into the
rotating frame of the Larmor precessing spin.

 The symmetry-breaking ferromagnetic domain formation discussed above is a phenomenon that  accompanies
a large number of thermodynamic phase transitions such as the paramagnetic to
 ferromagnetic phase transition
seen in iron when the temperature  is lowered below its Curie point.
However, unlike the thermodynamic phase
transition in Fe, the symmetry-breaking transition in the spinor BEC
is a quantum transition that can occur at arbitrarily low
temperature.  Concomitantly, the initial fluctuations that seed the
symmetry breaking dynamics in a spinor BEC need not be  thermal in nature
but, rather, may be of a quantum origin.
 The quantum origin of such fluctuations  makes the detailed study of the
origin and dynamics of the spontaneous magnetization  of fundamental interest.

One approach to developing a  theoretical understanding of the above  phenomenon is to analyze the
 low energy dynamics of a spinor BEC by linearizing the
Heisenberg equation of motion of the bosonic annihilation operators for the atoms.
In previous theoretical studies \cite{lamacraft,girvin}, these linearized equations of motion were
obtained  by expanding the fields corresponding
to the three components of the BEC around the initial $m_z=0$ state. One finds that
the low energy dynamics of the condensate in the initial state can be described by a set of three low energy excitations
 composed of a gapless phonon mode and two gapped magnon modes.
On quenching  the quadratic Zeeman shift to a value which is below  the phase transition,
the magnon modes are found to become unstable. These unstable modes amplify any initial perturbations
 from a homogeneous polar state. Thus  the domain formation following the quench is
described as resulting from the quantum fluctuations in the initial ground state being dynamically amplified\cite{lamacraft, girvin}.

The theoretical calculations discussed above treat the domain formation
by calculating the linearized dynamics of a homogeneous spinor condensate with
local interactions. While these calculations yield results in  qualitative
agreement with experiment, a quantitative comparison to experiment is
 essential for confirming the quantum nature of the
initial seed and of the amplifier driving the structure formation.
  An improved understanding of the dynamics of the spinor
condensate including the effect of the trapping potential and non-linearities
has been obtained  in previous theoretical works \cite{ueda_top,ueda_kz,girvin}.
In the current work, we  improve further our understanding of the dynamics
 of spinor
condensates by including the effects of dipole-dipole interactions
 and finite temperature and  use this
improved understanding to determine the domain formation resulting
from quantum fluctuations within the truncated Wigner approximation (TWA).

We begin with  a discussion of how  the condensate in a pancake-shaped trap can be modeled by a
two dimensional
Hamiltonian. Next we introduce a general framework for  calculating the eigenmodes of the
inhomogeneous gas which are used to describe the time evolution of the initial quantum fluctuations.
 Following this we discuss explicitly the effect of dipole-dipole
interactions, the trap and finite temperatures on the dynamics of a spinor BEC.  Next we introduce
 the approximations
and the general computational framework that allows us to include all these effects together with non-linearity effects
such as saturation of the transverse magnetization at long times.
 Finally we comment on how our results compare to experimental results.

\section{Two-dimensional effective Hamiltonian}
The many-body Hamiltonian for the spin-1 $^{87}$Rb gas considered in this work,
expressed using bosonic fields $\hat{\psi}_{m_p}$ to represent the three magnetic sublevels
$m_p=0,\pm 1$, is given as \cite{bigelow,ueda_top}
\begin{equation}
H=\int \,\ud\mathbf{R}\,\left\{\frac{\hbar^2}{2 M}\sum_\alpha|\nabla \hat{\psi}_\alpha|^2+c_0 (\hat{n}^2-\hat{n})+c_2 \hat{\mathbf{F}}^2+q(t) \hat{Q}+V_{trap}(\mathbf{R})\hat{n}\right\}+U_{dipole}
\end{equation}
where   $M$ is the atomic mass, and the quadratic Zeeman shift, $q(t)$ is applied along $\mathbf{P}$ and $V_{trap}(\mathbf{R})$ is the  external trap
potential. The parameter $c_0$ is the strength of  the spin-independent atom-atom short range repulsion,
   $c_2$ is the  coupling constant for the  spin-dependent contact interaction,
and $U_{dipole}$ is the  dipole-dipole interaction. The boson field operators appear in the magnetic part of the interaction terms through the spin density operators defined as
 $\hat{F}_{\alpha}=\sum_{\beta,\gamma}J_{\alpha,\beta,\gamma}\hat{\psi}_\beta^\dagger\hat{\psi}_\gamma$ where $J_{\alpha}$
 are the spin-1 matrices in the fundamental representation in a basis such that the  spin is quantized along $\mathbf{P}$, which is the axis determined by
the quadratic Zeeman shift. The quadratic Zeeman shift term in the Hamiltonian is of the form $\hat{Q}=\sum_{\beta,\gamma}(\mathbf{P}\cdot J)^2_{\beta,\gamma}\hat{\psi}_\beta^\dagger\hat{\psi}_\gamma$. The total atom density is given
by $\hat{n}=\sum_\alpha |\hat{\psi}_\alpha|^2$. In the above discussion,  the condensate is prepared in the
$m_p=0$ hyperfine state, where $m_p$ is the magnetic quantum number for spins quantized along the axis $\mathbf{P}$.
This state is prepared at a high quadratic Zeeman shift
$q(t)\gg 2|c_2|n_{3D}$ where $n_{3D}$ is the peak density
at the center of the condensate.  The quadratic Zeeman shift is then rapidly reduced below the critical value of  $2|c_2|n_{3D}$ and the state of the condensate
is allowed to evolve. In the experimental work discussed in this paper the axis along which the quadratic Zeeman shift is applied $\mathbf{P}$ is taken
to be the z-axis.

As mentioned in the introduction, the atoms in the system evolve under a strong magnetic field which we have eliminated from the Hamiltonian described in Eq.\ 1,
by transforming to a rotating frame. Such a transformation leaves the rotation-invariant terms in the Hamiltonian unchanged but affects the quadratic Zeeman
shift term and the dipole-dipole interaction term $U_{dipole}$. Yet, under the experimental conditions that the Larmor precession frequency
is far higher than that of the interaction-driven spin dynamics, one may consider the  quadratic Zeeman shift term and the dipole interaction as
precession-averaged static terms in the Hamiltonian. It may also be possible to  vary dynamically the axis of the quadratic Zeeman shift to follow the Larmor precession
of the initial state applied by the initial magnetic field, in which case the quadratic Zeeman shift term would be intrinsically stationary in the rotating
frame of the spin.

 The above defined Hamiltonian can be used to calculate the dynamics of the
transverse magnetization, which is the observed quantity that is used in the above mentioned
 experimental studies to observe
 the non-magnetic
to ferromagnetic phase transition\cite{dmsk_nature}. The $x$ and the $y$ components
of the transverse magnetization, $\hat{F}_x$ and $\hat{F}_y$, can be combined into a single complex transverse magnetization operator
$\hat{F}_{\perp}(\mathbf{R})=\hat{F}_x(\mathbf{R})+\hat{F}_y(\mathbf{R})$
which is given in terms of the fields $\hat{\psi}_\alpha$ by
\begin{equation}\label{eq:tm}
\hat{F}_{\perp}(\mathbf{R})=\sqrt{2}\left(\hat{\psi}_0^\dagger(\mathbf{R}) \hat{\psi}_{+1}(\mathbf{R})+\hat{\psi}_0(\mathbf{R})\hat{\psi}_{-1}^\dagger(\mathbf{R})\right)
\end{equation}
 The random magnetization domain pattern that forms after the quench can be characterized by a
correlation function of the above defined transverse magnetization, which we take as  $G(\delta\mathbf{R})=\int \,\ud \mathbf{R}\,\left\langle \hat{F}_\perp(\mathbf{R}-\delta\mathbf{R}/2) \hat{F}_\perp^\dagger (\mathbf{R}+\delta\mathbf{R}/2)\right\rangle/\left(\int \,\ud \mathbf{R}\, \langle \hat{n}(\mathbf{R})\rangle\right)^2$\cite{dmsk_nature}.

An important feature of the experiments is the use of condensates with widths in one dimension $(\hat{y})$ that are smaller than the spin healing length,
 $2\pi/\sqrt{2 M |c_2| n_{3D}}$.
 The three-dimensional Hamiltonian  is thus reduced to a two dimensional form through the substitutions
 $\hat{\psi}_\alpha(\mathbf{R})=\hat{\phi}_\alpha(\mathbf{r})\xi(\mathbf{r};y)\sqrt{n_{2D}}$ where $y=\mathbf{R}\cdot\hat{\mathbf{y}}$ and $\mathbf{R}=\mathbf{r}+y\hat{\mathbf{y}}$.
The peak density integrated along the $y$ axis is given by $n_{2D}=\int \,\ud y\,\left \langle \hat{n}(0,y)\right\rangle_{initial}$.
Here $\xi(\mathbf{r};y)$ represents the normalized spatial profile of the wavefunction   at each point $\mathbf{r}$  in the two dimensional
 $x$-$z$ plane.
The scale of the initial quantum zero-point fluctuations of the  two dimensional field, $\hat{\phi}_\alpha$, is determined
 from the canonical commutation relations $[\hat{\phi}_\alpha(\mathbf{r}),\hat{\phi}_\beta^{\dagger}(\mathbf{r'})]=\frac{1}{n_{2D}}\delta(\mathbf{r}-\mathbf{r'})\delta_{\alpha,\beta}$.

Using these relations and the Heisenberg equations of motion one can construct a time-evolution equation for the operator
 $\hat{\phi}_\alpha(\mathbf{r})$ of the form
\begin{align}
\imath\partial_t\hat{\phi}_\alpha(\mathbf{r})&=\left(-\frac{\hbar^2}{2 M}\right)\nabla_{\mathbf{r}}^2\hat{\phi}_\alpha(\mathbf{r})+\hat{\phi}_\alpha(\mathbf{r})\int \,\ud y \,\left(-\frac{\hbar^2}{2 m}\right)\left(\nabla^2\xi(\mathbf{r};y)\right)\xi(\mathbf{r};y)\nonumber\\
&+\hat{\phi}_\alpha(\mathbf{r})\int \,\ud y\,\xi^2(\mathbf{r};y)V_{trap}(\mathbf{r};y) +U_{interaction}
\end{align}
The second term on the right hand side of the above equation may be considered to be an effective renormalization of the potential energy
 related to confinement effects.
Within the Thomas-Fermi (TF) approximation this term is small through most of the condensate and is hence ignored.

For simplicity, we assume the wavefunction profile to be a TF profile given by $\xi(\mathbf{r};y)=\frac{3}{4 R_{\textrm{TF},y}(\mathbf{r})}(1-y^2/R_{\textrm{TF},y}^2(\mathbf{r}))^{1/2}$
where $R_{\textrm{TF},y}(\mathbf{r})=R_{\textrm{TF},y}(1-x^2/R_{\textrm{TF},x}^2-z^2/R_{\textrm{TF},z}^2)^{1/2}$ and $R_{\textrm{TF},x}$, $R_{\textrm{TF},y}$ and $R_{\textrm{TF},z}$ are the TF radii in the $x$, the $y$ and the $z$ directions respectively.
With the above choice of a transverse profile, the contribution of a local
 two-body potential of the form $c_m\delta(\mathbf{r}-\mathbf{r'})\delta(y-y')$
to the interaction term $U_{interaction}$ simplifies to
\begin{equation}
c_m\int \psi^3(\mathbf{R})\xi(\mathbf{r};y)\,\ud y= c_m
\frac{n_{3D}}{|\xi(\mathbf{0})|^2}\int \xi^4(\mathbf{r} ;y) \,\ud y\,\phi^3(\mathbf{r})=0.8\,  c_m n_{3D}\frac{R_{\textrm{TF},y}(\mathbf{0})}{R_{\textrm{TF},y}(\mathbf{r})}\phi^3(\mathbf{r})
\end{equation}
where the index $m$ is either 0 or 2 depending on whether we are referring to the spin independent or spin dependent parts of the contact
interaction, respectively. In the rest of the article we will be using a two-dimensional position-dependent effective
interaction by $c_m(\mathbf{r})=0.8\, c_m n_{3D}R_{\textrm{TF},y}(\mathbf{0})/R_{\textrm{TF},y}(\mathbf{r})$.

\section{Quantum Dynamics and Quantum Noise seeded domain formation}

Let us now describe the fluctuations and domain formation in terms of the two dimensional fields derived above.
 We consider the dynamics and low energy fluctuations of the initial $m_p=0$ state by shifting the
operator corresponding to the $m_p=0$ component by $\hat{\phi}_0=\sqrt{n(\mathbf{r})}+\hat{\eta}_0$, where
$n(\mathbf{r})=\langle\hat{\phi}_0^\dagger(\mathbf{r})\hat{\phi}_0(\mathbf{r})\rangle$ is the equilibrium density for the condensate
at large positive quadratic Zeeman shift of $q\gg |c_2|n_{3D}$.
The
Hamiltonian can now be expanded to second order in the small
fluctuations in the small fluctuation operators $\hat{\phi}_{\pm 1},\hat{\eta}_0$.  In this
Hamiltonian, the terms involving $\hat{\eta}_0$, which describe the scalar
Bogoliubov spectrum of phonons and free particles, separate from
those involving spin excitations; these latter terms provide the
following Hamiltonian
\begin{align}
H_{magnon}&=-\frac{\hbar^2}{2M}\int \hat{\phi}_{+1}^{\dagger}\nabla^2\hat{\phi}_{+1}+\hat{\phi}_{-1}^{\dagger}\nabla^2\hat{\phi}_{-1}\nonumber\\
&+\int \left(q(t)+c_2(\mathbf{r})n(\mathbf{r})+V_{trap}(\mathbf{r})+c_0(\mathbf{r}) n(\mathbf{r})\right)\left(\hat{\phi}_{+1}^{\dagger}\hat{\phi}_{+1}+\hat{\phi}_{-1}^{\dagger}\hat{\phi}_{-1}\right)\nonumber\\
&+\int c_2(\mathbf{r})n(\mathbf{r})\left(\hat{\phi}_{+1}\hat{\phi}_{-1}+\hat{\phi}_{+1}^\dagger\hat{\phi}_{-1}^\dagger\right).
\end{align}
For simplicity, the effect of the dipole-dipole interaction term has been ignored here and its discussion is postponed to Section V.
The dynamics of the magnetic degrees of
freedom obtained from the above approximate Hamiltonian  are given as
\begin{equation}
\partial_t \hat{\Phi}(\mathbf{r})=\left\{\imath  \left(-\frac{\hbar^2}{2 M}\nabla^2 +q(t) + V_{trap}(\mathbf{r})+(c_0(\mathbf{r})+c_2(\mathbf{r})) n(\mathbf{r}) \right)\sigma_z + c_2n(\mathbf{r}) \sigma_y \right\}\hat{\Phi}(\mathbf{r}).
\end{equation}
where we have introduced the spinor $\hat{\Phi}(\mathbf{r})=\left( \begin{array}{c}
  \hat{\phi}_{+1}(\mathbf{r})\\
\hat{\phi}_{-1}^\dagger(\mathbf{r})\\
\end{array}\right)$.

The above spinor equation of motion can be used to describe the dynamics of the condensate in terms of normal modes $\Upsilon_n^{(\pm 1)}(\mathbf{r})$ with frequencies
$\pm E_n$. The dynamics of $\hat{\Phi}(\mathbf{r},t)$ are then determined as
\begin{equation}\label{eq:expansion}
\hat{\Phi}(\mathbf{r},t)=\sum_{n,\sigma=\pm 1} \hat{d}_n^{(\sigma)} e^{\imath\sigma E_n t} \Upsilon^{(\sigma)}_n(\mathbf{r}).
\end{equation}
where
$\hat{d}_n^{(\sigma)}$
are the mode occupancy operators. The magnon
modes are stable when the eigenenergies $E_n$ are real, and unstable
when $E_n$ are complex.  It is these unstable modes that amplify
quantum fluctuations to generate macroscopic magnetization in the
quenched spinor gas.

In the case of a homogeneous condensate the normal modes $\Upsilon_n^{(\pm 1)}(\mathbf{r})$ can be reduced to
a product of a plane wave state and the momentum dependent two-component spinor that appears in standard treatments of the
linearized Gross-Pitaevskii equations. However the determination of these modes in the case of an inhomogeneous density
must be done in real space using explicit numerical diagonalization of a generalized eigenvalue problem.
In the case of a positive quadratic shift, which is the focus of in this article, the frequencies of these eigenmodes can
be shown to be either purely real or imaginary as discussed in Appendix A.
 A similar eignmode expansion  for a
trapped spinor condensate in the limit of vanishing quadratic Zeeman shift has been reported in previous work\cite{girvin}.

The quantum noise amplified by these unstable  modes is entirely contained in the
correlation function of the spinors relative to the initial state. Since the initial state of our system is assumed to be prepared
as a condensate of atoms in the $m_p=0$ state, the population in the $m_p=\pm 1$ states is negligible and the relevant correlator is given
 by $\left\langle \hat{\Phi}(\mathbf{r}_1,0)\hat{\Phi}^\dagger(\mathbf{r}_2,0)\right\rangle=\delta(\mathbf{r}_1-\mathbf{r}_2)(1+\sigma_z)/2$.
 In the description of the dynamics
of the condensate in terms of magnon modes, the quantum fluctuations become encoded in the quantum mode
occupancy operators $\hat{d}_n^{(\sigma)}$, which can be derived from the spinor operator at the
time of quench  $\hat{\Phi}(\mathbf{r},0)$, $\hat{d}_n^{(\sigma)}=\int \,\ud\mathbf{r}\, \tilde{\Upsilon}_n^{(\sigma)*}(\mathbf{r})\hat{\Phi}(\mathbf{r},0)$, where $\tilde{\Upsilon}_n^{(\sigma)}(\mathbf{r})$ are the dual modes,
 explicit
expressions for which can be found in Appendix A.

Given this initial noise and the linear dynamics of these magnon modes, we may calculate the magnetization correlations that may be observed at short times
after the quench. Linearizing the transverse magnetization as
 $\hat{F}_\perp(\mathbf{r})=n_{2D}\sqrt{2 n(\mathbf{r})}\left(\hat{\phi}_{+1}^\dagger(\mathbf{r})+\hat{\phi}_{-1}(\mathbf{r})\right)$, we obtain
\begin{align}
&G(\delta\mathbf{r})=\frac{\int \,\ud \mathbf{r}\langle \hat{F}_\perp(\mathbf{r}-\delta\mathbf{r}/2)\hat{F}_\perp^\dagger(\mathbf{r}+\delta\mathbf{r}/2) \rangle}{\left(\int \,\ud\mathbf{r}\, n_{2D}n(\mathbf{r})\right)^2}=
\sum_{n,m,\sigma,\sigma'}e^{\imath(\sigma E_n-\sigma'E_m^*)t}\nonumber\\
&\times\frac{\int \,\ud \mathbf{r}\sqrt{n(\mathbf{r}-\delta\mathbf{r}/2)n(\mathbf{r}+\delta\mathbf{r}/2)}\,\Upsilon_m^{(\sigma')\dagger}(\mathbf{r}-\delta\mathbf{r}/2)(\mathbf{1}+\sigma_x)\Upsilon_n^{(\sigma)}(\mathbf{r}+\delta\mathbf{r}/2)}{\left(\int \,\ud \mathbf{r}\,n(\mathbf{r})\right)^2}\nonumber\\
&\times\int \,\ud\mathbf{r}_1\,\tilde{\Upsilon}_n^{(\sigma)\dagger}(\mathbf{r}_1)(\mathbf{1}+\sigma_z)\tilde{\Upsilon}_m^{(\sigma')}(\mathbf{r}_1)
\end{align}

Note that the  correlation function  defined above suffers from a UV  divergence.
The physical origin of this divergence is the fact that $\hat{F}_{\perp}$ represents the magnetization of  point-like
particles.
This UV divergence is however not observed experimentally because of the physically natural cut-offs like the finite spatial and  temporal resolution
of the measuring apparatus, which introduces a natural spatio-temporal averaging into the observed magnetization. For our
purposes it suffices to consider only the contribution of unstable magnon modes to the transverse magnetization.
 Hence in our calculations we avoid this UV divergence in the correlator of $\hat{F}_\perp$ by restricting the above sum
to  imaginary frequency modes.

\section{Non-linearity effects: Truncated Wigner Approximation}

In the last section we saw how we can describe the physics of the quench by expanding the Heisenberg equations of motion
about the initial condensate state and keeping terms up to linear order in the fluctuations. Even though this might be
expected to be a relatively accurate description at short times where deviations from the initial state are small, it
 breaks
down at longer times and predicts an unphysical diverging magnetization.

Such a divergence is
avoided by considering the complete Hamiltonian, including the
higher order terms neglected in our prior approximation.  The direct
solution of the Heisenberg equations of motion with the non-linearity terms would present an extremely difficult task.
This is a general feature of problems involving  quantum many-particle systems and for this
reason various approximations must be used to understand such problems.
One such approximation, that has been seen to describe the dynamics of BECs reasonably well\cite{haine},
 is the  Truncated Wigner Approximation (TWA)\cite{gardiner,gardiner_book}.

Within the TWA one is interested in calculating the time evolution of the
Wigner Distribution Function (WDF) of the fields  $\hat{\phi}_\sigma(\mathbf{r})$, which are the quantum analog of the classical phase space distribution.
The fields are assumed to evolve according to the Gross-Pitaevskii equation (GPE) which is the mean field version of the
Heisenberg equation of motion where the operator $\hat{\phi}_\sigma(\mathbf{r})$ has been replaced by the time-dependent order parameter $\phi_\sigma(\mathbf{r})$.
Quantum fluctuations around the mean field time evolution are included in the TWA by adding to the initial condition of the
order parameter a noise term  which is picked randomly from a classical distribution of fields corresponding
to the initial WDF. In our case, where the initial state has a negligible
population in the $m_p=\pm 1$ hyperfine state,
the classical distribution of the randomly picked wavefunctions, $\phi_\sigma(\mathbf{r})$,  is a Gaussian distribution
 with variance given by
 $\langle \phi_{\sigma}^*(\mathbf{r})\phi_{\sigma'}(\mathbf{r}')\rangle_{classical}=\frac{1}{2}\langle \Psi_{initial}|\{\hat{\phi}_{\sigma}^\dagger(\mathbf{r})\hat{\phi}_{\sigma'}(\mathbf{r}')\}|\Psi_{initial}\rangle$ where $|\Psi_{initial}\rangle$ is the initial state with all atoms in the $m_p=0$ hyperfine state.
\footnote{To see this we observe that the requirement of equality  of the quantum and the classical distributions
 is equivalent to the equality of the quantum characteristic function
 $\chi_W(\lambda,\lambda^*)=\langle \exp(\hat{\phi}^\dagger\lambda-\lambda^\dagger\hat{\phi})\rangle
 =\prod_n\langle \exp(\hat{c}^\dagger_n(u_n^\dagger\lambda-\lambda^\dagger v_n^*)+\hat{c}_n(v_n^T\lambda-\lambda^\dagger u_n))\rangle$
to the classical characteristic function $\chi_{classical}(\lambda,\lambda^*)=\langle \exp(\phi^\dagger\lambda-\lambda^\dagger\phi)\rangle
=\prod_n\langle \exp(c_{n}^\dagger(u_n^\dagger\lambda-\lambda^\dagger v_n^*)+ c_{n}(v_n^T\lambda-\lambda^\dagger u_n))\rangle$.
Thus the 2 characteristic functions agree if
$\langle \exp(\hat{c}^\dagger_n\gamma+\hat{c}_n\gamma^*)\rangle=\langle \exp(c_{n}^*\gamma+ \gamma^* c_{n})\rangle=\frac{2}{\pi}\exp(-2|\gamma|^2)$
}.

\begin{figure}
\centering
\includegraphics[scale=0.4]{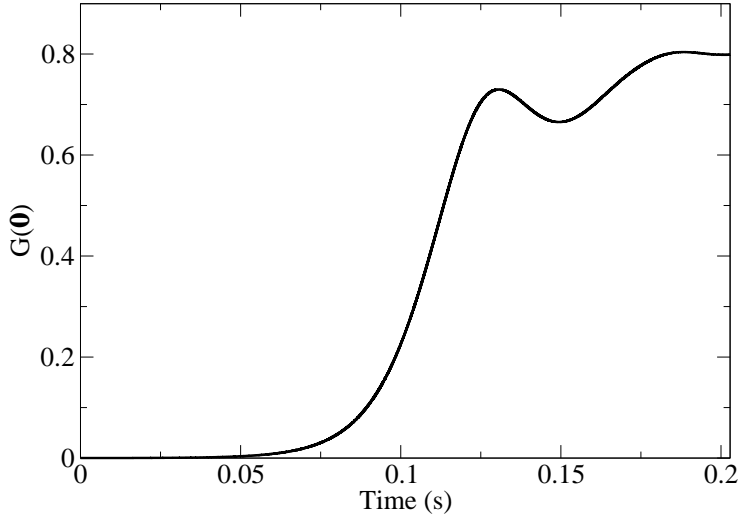}
\caption{Evolution of the variance of the transverse magnetization for the experimental configuration\cite{dmsk_prl} for $q/h=2$ Hz  and $c_2(\mathbf{0})/h=8$ Hz, calculated using the
TWA to include non-linearity induced saturation effects.}
\end{figure}

Within the TWA, we obtain the magnetization correlation
function by averaging over numerical results obtained for the
different, random representations of quantum noise.  As shown in
Fig.\ 1, this procedure yields a satisfactory result for the
magnetization variance $G(\mathbf{0})$ that saturates rather than
diverging.

The TWA has been shown to describe dynamical phenomena in BECs with reasonable success.
However the TWA  fails to describe certain aspects of the dynamics of BECs, such as the non-condensate fraction,
 and improvements beyond the TWA
have been proposed in several works\cite{polkovnikov,drummond}. It is difficult to estimate the accuracy of the TWA in
 the full multi-mode spinor condensate system that we are
studying. However as has been observed in previous work \cite{bigelow},
 it is possible to solve the Hamiltonian under consideration within
the single mode approximation exactly. This corresponds to the limit of a small trap where we can ignore the spatial dependence of the dynamics
of the atoms completely and the system of $2N$ atoms can be described by a Hamiltonian given by $\hat{H}_{SMA}=(2N-\hat{\phi}_0^\dagger\hat{\phi}_0)(\hat{\phi}_{+1}^\dagger\hat{\phi}_{+1}+\hat{\phi}_{-1}^\dagger\hat{\phi}_{-1})-(\hat{\phi}_{+1}^\dagger\hat{\phi}_{-1}^\dagger\hat{\phi}_0^2+\hat{\phi}_0^{\dagger 2}\hat{\phi}_{+1}\hat{\phi}_{-1})$. The time evolution of the fluctuations in the transverse magnetization of this Hamiltonian
can be determined exactly by numerically solving the time evolution of an initial state where all atoms are in the $m_p=0$ hyperfine state.
 We compared the results of this calculation for a system of 2000 atoms
to the time evolution of the transverse magnetization obtained within the TWA
for the same system and found excellent agreement between the growth rate within the TWA to exact results up to the saturation time within the TWA.
The exact transverse magnetization was found to increase past the saturation value within the TWA to a value that was 10\% higher than the TWA saturation.
 While it is  possible that the  multimode non-linearity of our system causes
 physics beyond the TWA to become directly relevant, the above comparisons of the TWA to exact results for the single mode systems demonstrate that the TWA accounts
for some of the effects of non-linearity in these systems.

\section{Finite temperature effects from an initial phonon population}

In attempting to make quantitative comparisons between calculations and
experimental observations, it is imperative to consider the role of the
non-zero temperature on the initial preparation and later evolution of the
experimental system. In fact, at first glance, one might expect the quench experiments
reported to be wholly dominated by thermal effects, given that the gas is
prepared by evaporative cooling at a temperature of $T\approx 50$ nK for which the
thermal energy is far larger
than the spin-dependent energies responsible for the quench dynamics, i.e.\
$k_B T\gg|c_2|n_{3D}$. However, one must consider separately the \emph{kinetic}
and \emph{spin} temperatures of the paramagnetic condensate in these
experiments. While the thermal population of the scalar excitations, the
$m_p=0$ Bogoliubov excitations about the $m_p=0$ condensate is indeed
determined by the $T=50$ nK kinetic temperature of the gas, the magnon
excitations are expelled from the gas by the application of magnetic field
gradients that purify the $m_p=0$ atomic population. To the extent
that such state purification is effective, and that magnon excitations are not
thermally produced, e.g. by incoherent spin-exchange collisions, the initial
\emph{spin} temperature of the system is indeed near zero. Thus remarkably,
a purely
quantum evolution may indeed occur in the non-zero temperature gas.

Here, we consider the possible influence of the thermal population of
scalar excitations on the quantum quench experiments.
The study of the coupling of phonons and magnons requires going beyond the linearized Heisenberg equation of motion. Thus, we consider the time
evolution operator for the quantum state of the spinor condensate as a coherent state path integral $U(t_1,t_2)=\int \prod_\alpha D\phi_\alpha D\phi^*_\alpha \exp(\imath S[\phi_\alpha,\phi^*_\alpha])$, as has been found useful
for many boson problems\cite{sachdev}. Here $S$ is the action for the three-component boson field corresponding to the Hamiltonian in Section II.
The scalar  phonon fluctuations are composed of a scalar density fluctuation,  $\delta n(\mathbf{r},t)=\sum_\alpha|\phi_{\alpha}(\mathbf{r},t)|^2-n(\mathbf{r})$, and current fluctuations associated with the density fluctuations,
required
by number conservation. In the case of a condensate with population
 dominantly in the $m_z=0$ hyperfine state, the current fluctuations can be described by
the superfluid phase $\lambda({\bf{r}}, t)$ defined through
$\phi_0(\mathbf{r},t)=e^{\imath \lambda(\mathbf{r},t)}\sqrt{n(\mathbf{r})+\delta n(\mathbf{r},t)-|\phi_{+1}(\mathbf{r},t)|^2-|\phi_{-1}(\mathbf{r},t)|^2}$.
 Since the density fluctuations are gapped at a high energy by the term $c_0 \delta n(\mathbf{r},t)^2/2$ in the
 action $S$,  they can be integrated out to leave an effective action
involving the phase $\lambda(\mathbf{r},t)$. Therefore in order to eliminate the high frequency density  fluctuations we perform the field substitution
 $\phi_{\pm 1}(\mathbf{r},t)\rightarrow \phi_{\pm 1}(\mathbf{r},t) e^{\imath\lambda(\mathbf{r},t)}$
in the action $S$ and then integrate out  the density fluctuations $\delta n(\mathbf{r},t)$.
This leads to the approximate action $S_{approx}=S_{phonon}[\lambda]+S_{magnon}[\phi_{\pm 1}]+V_{coupling}[\phi_{\pm 1},\lambda]$ where
the  $S_{phonon}=n\dot{\lambda}^2/2 c_0+n\frac{\hbar^2}{2 M}(\nabla \lambda)^2$, $S_{magnon}$ is the usual action for the $\phi_{\pm 1}$ atoms without the scalar interaction term
and $V_{coupling}=\frac{\hbar^2}{2 M}[-\nabla \lambda\cdot\sum_{\alpha}\textrm{Im}(\phi_\alpha\nabla\phi_\alpha^*)]$. The last term is the interaction that describes the
coupling between phonons and magnons.

The non-zero kinetic
temperature of the gas gives causes (low frequency) fluctuations of
superfluid phase with variance given by $\langle|\lambda_k|^2\rangle=2 M k_B T/\hbar^2 k^2 n_{2D}$.
 These thermal phase fluctuations couple to the dynamics of the magnons, through the
interaction term $V_{coupling}$.  A rough estimate  of the magnitude of the effect of the kinetic temperature
 can be  made  by considering the dimensionless ratio of the r.m.s value of  $V_{coupling}$ to the spin mixing energy $|c_2|n$ which
  is given by $\sqrt{\hbar^2 k_{spin}^2/2 M}\sqrt{(k_B T)k_{phonon}^2/ n_{2D}}/ |c_2| n_{3D}\approx \sqrt{(2 M k_B T)|c_2|/\hbar^2 n_{2D}c_0}$ where $k_{spin}$ is the wavevector associated with the spin healing length
and $k_{phonon}$ is the small wavevector associated with a phonon at the energy scale of the spin dynamics, $|c_2|n_{3D}$.
 For the kinetic temperature in experiment of 50 nK, this dimensionless parameter characterizing thermal effects is found to be less than $1.3\times10^{-2}$. A more rigorous
calculation within the TWA, where phonons are introduced by adding random thermal fluctuations to the initial conditions in $\phi_0$, confirms our rough estimate by showing
a negligible effect of the kinetic
temperature.

\section{Role of Dipole-Dipole interactions}

In the preceding paragraphs we have discussed the physics of the formation of domains from
 quantum fluctuations in a trapped quasi-two-dimensional condensate with ferromagnetic interactions.
However theoretical\cite{doniach,whitehead} and experimental studies\cite{dmsk_dipole}
 suggest that dipolar interactions play an important role in determining the magnetization textures for this system.
In this section, we provide the first
characterization of the role of dipolar interactions on the quantum quench
dynamics of a $^{87}$Rb spinor BEC.

The atomic spin undergoes Larmor precession at a high frequency, on the order of tens
of kHz, even as slower dynamics responsible for spontaneous magnetization transpire.
While this Larmor precession has no influence on average on the spin dependent s-wave contact
 interaction or the quadratic Zeeman shift, the time averaged Larmor precession of the atoms must be accounted for in calculating the
 influence dipolar interactions,
yielding an effective precession-averaged interaction of the form\cite{ueda_dip,demler_dipole}
\begin{align}
U_{dipole}&=\frac{\mu_0}{8 \pi}(g_F \mu_B)^2\int \,\ud \mathbf{R}_1\,\ud \mathbf{R}_2\,\frac{(\mathbf{R}_1-\mathbf{R}_2)^2-3(\mathbf{D}\cdot(\mathbf{R}_1-\mathbf{R}_2))^2}{|\mathbf{R}_1-\mathbf{R}_2|^5}\nonumber\\
&[3(\mathbf{D}\cdot\hat{\mathbf{F}}(\mathbf{R}_1))(\mathbf{D}\cdot\hat{\mathbf{F}}(\mathbf{R}_2))-\hat{\mathbf{F}}(\mathbf{R}_1)\cdot\hat{\mathbf{F}}(\mathbf{R}_2)]
\end{align}
where $\mathbf{D}$ is the dipole-precession axis (the magnetic field axis), $g_F=1/2$ is the gyromagnetic ratio of the electron,
$\mu_B$ is the Bohr magneton.
Integrating over the thin dimension of the condensate, we derive an effective two-dimensional
dipole interaction as
\begin{align}
U_{dipole}&= \frac{c_{dd}}{2}\int \,\ud \mathbf{r}_1\,\ud
\mathbf{r}_2\,K(\mathbf{r_1},\mathbf{r_2})[3(\mathbf{D}\cdot\hat{\mathbf{F}}(\mathbf{r}_1))(\mathbf{D}\cdot\hat{\mathbf{F}}(\mathbf{r}_2))-\hat{\mathbf{F}}(\mathbf{r}_1)\cdot\hat{\mathbf{F}}(\mathbf{r}_2)]\\
K(\mathbf{r},\mathbf{r}')&=\frac{1}{\xi^2(\mathbf{0})}\int
\frac{(\mathbf{R}-\mathbf{R}')^2-3(\mathbf{D}\cdot(\mathbf{R}-\mathbf{R}'))^2}{|\mathbf{R}-\mathbf{R}'|^5}\xi^2(\mathbf{r};y)\xi^2(\mathbf{r}';y')\,\ud y \,\ud y'
\end{align}
where the dipole interaction strength is given by $c_{dd}=\frac{n_{3D}\mu_0 }{4 \pi}(g_F \mu_B)^2$.
The dipole-dipole interaction term $U_{dipole}$ is a spin-dependent  interaction term in the Hamiltonian in addition to the ferromagnetic
part of the contact interaction  already discussed. The total of the two spin-dependent parts of the interaction Hamiltonian is given by
\begin{align}
H_{spin} &\equiv  \int \,\ud\mathbf{r}\,c_2(\mathbf{r}) \hat{\mathbf{F}}(\mathbf{r})^2+U_{dipole}=\int \,\ud \mathbf{r}_1\,\ud \mathbf{r}_2\,c_2^{\mbox{\scriptsize{\mbox{\scriptsize{eff}}}}}(\mathbf{r_1},\mathbf{r_2})\frac{[\hat{\mathbf{F}}(\mathbf{r}_1)\cdot\hat{\mathbf{F}}(\mathbf{r}_2)]}{\sqrt{n(\mathbf{r}_1)n(\mathbf{r}_2)}}\nonumber\\
&+\frac{3}{2}\int\,\ud \mathbf{r}_1\,\ud
\mathbf{r}_2\,c_{dd}K(\mathbf{r_1},\mathbf{r_2})(\mathbf{D}\cdot\hat{\mathbf{F}}(\mathbf{r}_1))(\mathbf{D}\cdot\hat{\mathbf{F}}(\mathbf{r}_2))\\
c_2^{\mbox{\scriptsize{eff}}}(\mathbf{r_1},\mathbf{r_2})&=c_2(\mathbf{r_1}) n(\mathbf{r_1})\delta(\mathbf{r_1}-\mathbf{r_2})-c_{dd}\sqrt{n(\mathbf{r}_1)n(\mathbf{r}_2)}K(\mathbf{r}_1,\mathbf{r}_2)/2
\end{align}
Thus apart from renormalizing the spin-dependent part of the contact interaction to $c_2^{\mbox{\scriptsize{eff}}}$, the dipole-dipole interaction also has an intrinsically anisotropic
contribution which is given by the second term in Eq.\ 12, where the anisotropy is not related to the spatial anisotropy of the dipole interaction kernel $K$.
 This term however turns out to not be relevant for the linearized dynamics in the case where the dipole precession axis $\mathbf{D}$
coincides with the spin-quantization axis $\mathbf{P}$.

 In the homogeneous case $c_{2}^{\mbox{\scriptsize{eff}}}$ can be written as below
\begin{equation}\label{eq:dipc2}
c_2^{\mbox{\scriptsize{eff}}}(k,\chi,\eta)=c_2(\mathbf{0})-\frac{c_{dd}}{2}K(k,\chi,\eta)
\end{equation}
where $K(k,\chi,\eta)$ is the dipole interaction kernel,
and  $\chi$ is the angle that $\mathbf{D}$ makes with the $y$-axis and
$\eta$ is the polar angle of the vector $\mathbf{D}$ in the plane of the BEC, as shown in Fig.\ 2.
 The wave-vector $k$ is taken to be along the $z$-axis
in the plane of the BEC.
As discussed in Appendix A, the eigenmode treatment discussed in Section III can be easily generalized to include dipole-dipole interactions.

From Eq.\ \ref{eq:dipc2} it is apparent that in the three-dimensional homogeneous case dipole-dipole
 interactions enhance structure formation
for wave-vectors  along
 the dipole-precession axis and suppress it for wave-vectors transverse to the dipole-precession axis.
However the effect of dipole-dipole interactions on a quasi-two-dimensional condensate is qualitatively
different.
The Fourier transform of the interaction $K(k,\chi,\eta)$ in the case of the parabolic TF transverse
 profile
along the $y$ direction is difficult to compute analytically. To obtain a qualitative understanding
 of dipole interactions we consider the case of a 2D condensate
for the case of
a Gaussian profile $\xi(\mathbf{r};y)$ of width $w=\frac{4}{3\sqrt{2\pi}}R_{\textrm{TF},y}$. This width is chosen so that the peak density for the normalized profile matches
 that of the TF profile.
 The expression used for the Gaussian regularized dipole interaction derived in the Appendix B can be used in conjunction with standard integrals
 to determine the Fourier transform of $K(\mathbf{r})$ for this Gaussian choice of profile to be
\begin{equation}\label{eq:F}
K(k,\chi,\eta)=\sqrt{2 \pi}\left[\frac{ 4 \sqrt{\pi}}{3 }(1-\frac{3}{2}\sin^2\chi)+\pi k w(\sin^2\chi \cos 2 \eta+3 \sin^2 \chi-2)(1-\textrm{Erf}(k w))e^{k^2 w^2}\right].
\end{equation}
\begin{figure}
\centering
\includegraphics[scale=0.5]{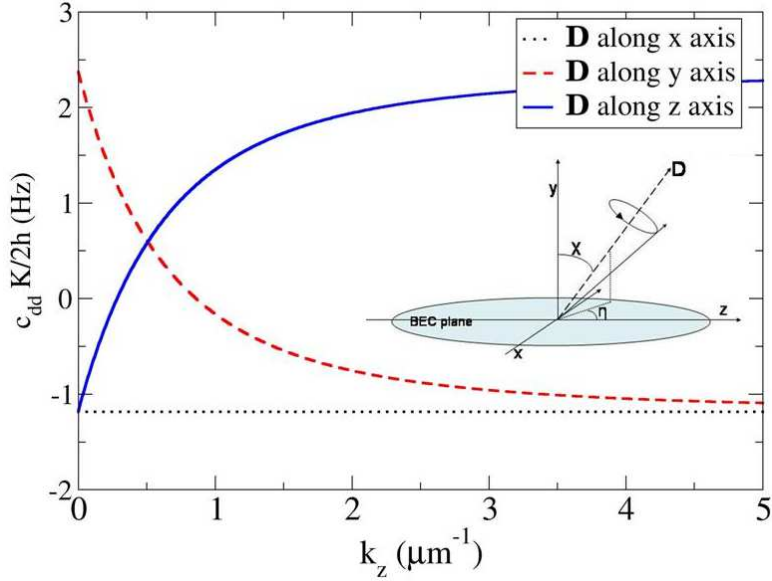}
\caption{Momentum dependence of the contribution of the dipole interaction kernel K defined in Eq.\ \ref{eq:F} to $c_{2}^{\mbox{\scriptsize{eff}}}$ for $c_{dd}/h=0.8$ Hz, $c_2(\mathbf{0})/h=-8.0$ Hz.
 The inset shows the orientation of the dipole-precession axis  $\mathbf{D}$ relative to
the coordinates and the plane of the BEC.}
\end{figure}
The contribution of the momentum variation of the dipole interaction kernel $K$ to $c_{2}^{\mbox{\scriptsize{eff}}}$  is shown in Fig.\ 2.
In the large $k$ limit, this expression, apart from a factor of $\sqrt{2}$ arising from the re-normalization because of the
transverse profile, reduces to   $-\frac{4 \pi}{3}(1-3 \sin^2\chi \cos^2\eta)$ which is the three dimensional form as expected. As seen in Fig.\ 2, the small $k$ limit is found to be consistent
with previous theoretical studies\cite{demler_dipole}.

In the case where the dipole-precession axis $\mathbf{D}$ coincides with the spin-quantization axis of the atoms $\mathbf{P}$,
 we can use the explicit form for $c_{2}^{\mbox{\scriptsize{eff}}}$ given in Eq.\ \ref{eq:dipc2} to discern  the effect of the anisotropic
dipole-dipole interactions on behavior of the spin dynamics
  by studying the dispersion relation $E^2(\mathbf{k})=(\frac{\hbar k^2}{2 M}+q)(\frac{\hbar k^2}{2 M}+q+2
c_{2}^{\mbox{\scriptsize{eff}}}(\mathbf{k}))$ in the presence of
dipole-dipole interactions.
As seen in Fig.\ 3,  when the dipole-precession axis points along the long axis of the condensate, i.e
the $z$-axis, as in the experiments, the effect of the dipole interaction is weak and the dipole interactions
 slightly shorten the length
 scale and lengthen the time scale of domain formation.
In contrast, the domain formation is significantly slowed down by the
 dipole-dipole interaction when $\mathbf{D}$ is oriented along the
$x$-direction. Interestingly
when the dipole axis is pointed along the $y$-axis, the thin axis  of the condensate, the rate of structure
formation is dramatically increased in both directions.

\begin{figure}
\centering
\includegraphics[scale=0.4]{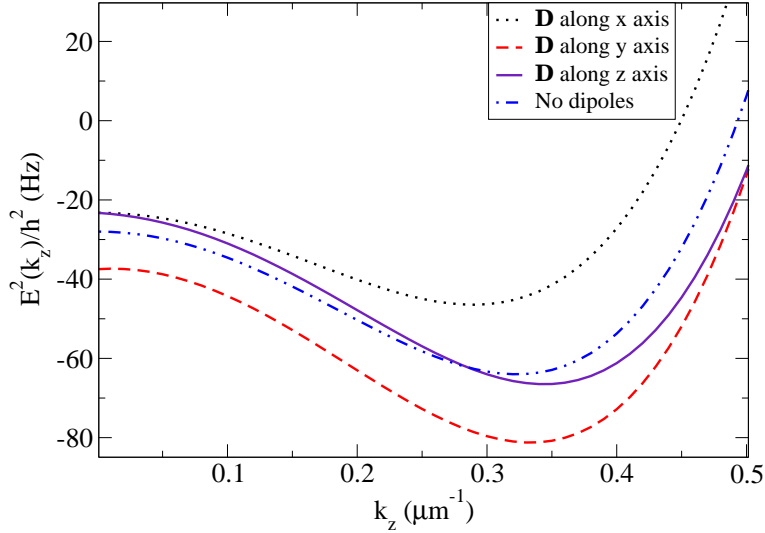}
\caption{Magnon dispersion curves for an unbounded two-dimensional condensate in the $x-z$ plane, for $q/h=2$ Hz, $c_2(\mathbf{0})/h=8$ Hz and $c_{dd}/h=0.8$ Hz, including the effects of
 dipole-dipole interaction for $\mathbf{D}$ aligned along $\hat{\mathbf{x}}$,$\hat{\mathbf{y}}$ or $\hat{\mathbf{z}}$, and the
wave-vector being assumed to be aligned along $\hat{\mathbf{z}}$. }
\end{figure}

The effects of
the anisotropic dipolar interactions may also be highlighted in
quantum quenches where the dipole-precession axis $\mathbf{D}$ differs from the spin-quantization axis $\mathbf{P}$.
Specifically, consider  the case where $\mathbf{D} = \hat{y}$, while $\mathbf{P}$ is
prepared to be orthogonal to $\mathbf{D}$ (i.e.\ the axis $\mathbf{P}$ Larmor
precesses in the $\hat{x}-\hat{z}$ plane). In this case the intrinsic spin-anisotropic term
in Eq.\ 12 can no longer be ignored when constructing the linearized dynamics of a two-dimensional homogeneous condensate and this leads to a contribution which breaks the symmetry of the two polarizations of the magnon modes describing the magnetization
dynamics of the condensate.
Altogether the  strong variation of post-quench dynamics with changes in the system geometry provides a compelling
signature of dipole-dipole interactions that may be studied in  future experiments.

\section{Numerical Methods and Results}

Having set up our theoretical model we now turn to the numerical techniques and quantitative results based on these ideas applied to a model spinor condensate
with parameters motivated from experiment.
As previously discussed, the calculation of correlation functions within the TWA requires the time evolution of an initial state
which is comprised of an initial  mean field state with random fluctuations added to it.
 The initial wavefunction of the $m_z=0$ condensate is determined by minimizing the total energy via
conjugate gradient minimization assuming $q\rightarrow\infty$. Time evolution according to the  GPE is determined numerically  by the 6-th order
 Runge Kutta method\cite{numerical_recipes} with periodic boundary conditions in space.  The kinetic energy is computed by Fourier transforming each component into momentum
space.
 The dipole-dipole interaction kernel, $K(\mathbf{r},\mathbf{r}')$,  has the properties both of being long-ranged and also of being  singular at
short distances. Therefore it is necessary to  regularize and truncate $K(\mathbf{r},\mathbf{r}')$ in real space before calculations are
 performed in Fourier space  to avoid interaction between inter-supercell periodic images as discussed in Appendix B.
In calculating $K(\mathbf{r},\mathbf{r}')$ for use in the solution of the full GPE, we neglect the variation of the condensate thickness $(R_{TF,y})$  along the $\hat{x}$ direction.
We have checked that this approximation doesn't significantly affect our results when $\mathbf{D}$ is along the long axis of the trap i.e. $\hat{z}$,
as is the case in experiment.

 For the calculations reported we use a time step of $3.5\,\mu$s and a grid spacing of $0.5\,\mu$m and the results are found to be converged
with respect to these parameters. In addition,
 the total energy of the system remains conserved to a certain error tolerance in the time evolution. It is also verified that
the total magnetization along $\mathbf{D}$ is a conserved quantity  in the absence of dipole-dipole interactions.

 The trap geometry for our calculations  is taken to be similar to experiment\cite{dmsk_prl} such that the TF radii of the condensate are $R_{\textrm{TF},x}=20\,\mu \textrm{m},R_{\textrm{TF},y}=1.6\,\mu \textrm{m}$. Given that the relevant lengthscale for spin dynamics $(2\pi/\sqrt{2 M |c_2| n_{3D}}\approx 2 \mu \textrm{m})$ is much smaller than the
$\hat{z}$ length of the condensate $(R_{TF,z}=200\,\mu\textrm{m})$, here we treat the system as unconfined along $\hat{z}$, with periodic boundary conditions over a 90 $\mu \textrm{m}$
length.
The peak three-dimensional and two-dimensional  densities are taken to be  $n_{3D}=2.5\times 10^{14}/\textrm{cm}^3$ respectively. The strength of the spin-dependent part of the contact interaction has been
inferred previously from molecular spectroscopy \cite{greene,verhaar} and from spin-mixing dynamics \cite{bloch,chapman}. According to these works $c_2(\mathbf{r}=0)=0.8\,|c_2|n_{3D}$
is predicted to lie between $h \times 6$ Hz and $h \times 8$ Hz, corresponding to $1.1\, a_B < \Delta a=(a_0-a_2) < 1.9 \,a_B$ where $\Delta a$ is the difference between the $s$-wave scattering
lengths for the spin-0
and spin-2 channels and $a_B$ is the Bohr radius.
  This variation  in the rate of spin-amplification makes it difficult to estimate how close the estimated initial noise from experiment is to
 the quantum limit.

 We find our results to be in qualitative agreement with experiment and previous theoretical calculations.
In particular, we find that  the average magnitude of the transverse magnetization grows exponentially from a small value
to a much larger value (Fig.\ 1) with a time-constant
that is relatively insensitive to the quadratic Zeeman shift $q$.
 The calculated domain structure and magnetization correlations match with
those observed experimentally, and in previous calculations\cite{ueda_kz},
 the characteristic domain size increasing with $q$ (Fig. \ 4). However as seen from Fig.\ 5 our calculations
somewhat underestimate the domain size for the larger of the experimentally measured values of $\Delta a$.
This discrepancy between theory and experiment is reduced on using the smaller of the measured values of
$\Delta a$. Thus the difference between theory and experiment could be the result of an error in the
 experimentally measured value of the
spin-dependent contact interaction or quantum and thermal effects of interactions that are not contained in
 the Gross-Pitaevskii equations. The introduction of a dipole-dipole
 interaction introduces a weak
dependence of the average local transverse magnetization $G(\mathbf{0})$ on the quadratic Zeeman shift.

\begin{figure}
\centering
\includegraphics[scale=0.4]{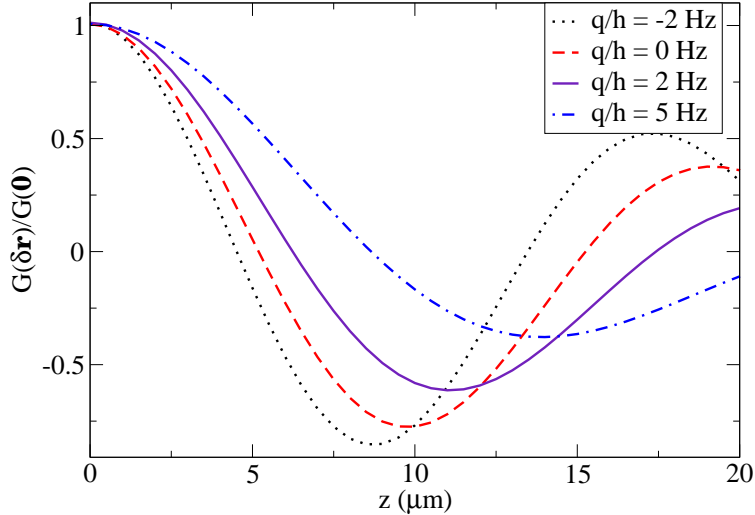}
\caption{Correlation function $G(\delta\mathbf{r}=z\hat{\mathbf{z}})$ at $t=100$ ms for a spinor condensate calculated using the method and geometry described in Section VII for various quadratic Zeeman shifts. The
correlation function plotted along the length of the condensate shows decreasing
 domain size with decreasing quadratic Zeeman shift.}
\end{figure}

\begin{figure}
\centering
\includegraphics[scale=0.5]{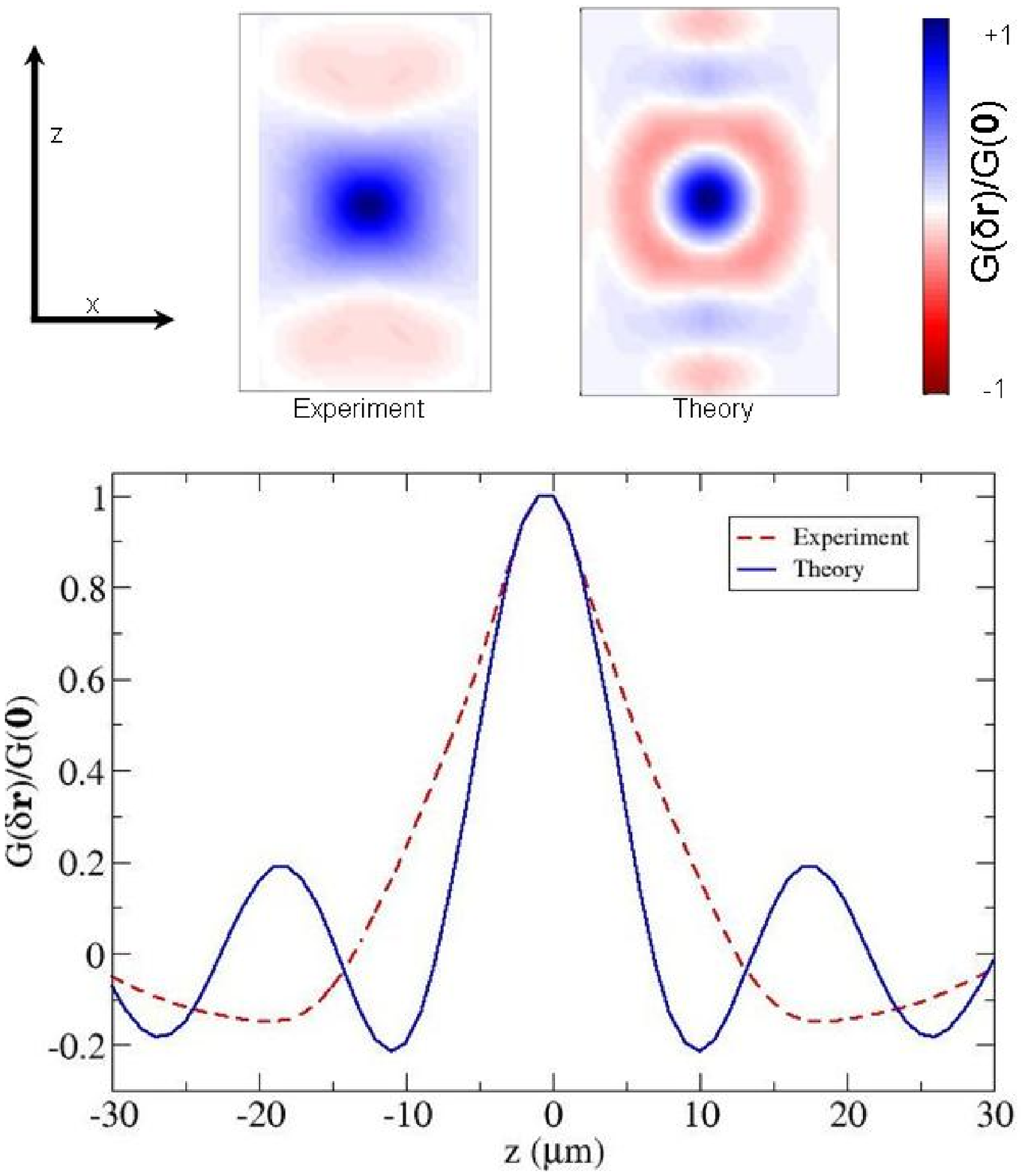}
\caption{Planar correlation function $G(\delta\mathbf{r})$ at $t=87$ ms
for a spinor condensate calculated using the method and geometry described in Section VII for $q/h=2$ Hz. The
one dimensional plots shown in the lower half represent sections of the two dimensional plots above through the center of the condensate at $x=0$. The two dimensional plots are 100 $\mu m$ along the
$z$ direction and 40 $\mu m$ along the $x$ direction. The above plots compare the calculated correlation structure  on the left to the experimentally measured structure on the right\cite{dmsk_prl}.}
\end{figure}

Despite the qualitative agreement between the homogeneous 2D condensate calculation \cite{lamacraft} and the current results, we find
quantitative differences between the results of the homogeneous case and the calculations including the trap and dipole interactions
 that are important for comparison to experiment. We discuss several of these differences below.

\subsection{Effect of the trapping potential}
Similar to previous theoretical work \cite{girvin}, we find a significant effect of the inclusion of the trap on the spin
dynamics in the parameter regime corresponding to experiment.
The external trapping potential along the width of the condensate, which is accounted for in our numerical calculations,
is found to slow  the growth of the transverse magnetization in the condensate significantly as seen in Fig.\ 7,
 when compared
to the quasi-two-dimensional homogeneous case without a trapping potential along the width.
 This slowing down can be understood intuitively
 from the
fact that the trap causes the density away from the center of the trap to be lower than at the center of the trap.
 Consistent with previous theoretical work\cite{girvin}, the density reduction away from the center of the trap
 also affects the spatial
structure of the correlations observed and the trap is found to  suppress the formation  of structure
in
the radial direction, as seen in Fig.\ 5.

\begin{figure}
\centering
\includegraphics[scale=0.35,angle=0]{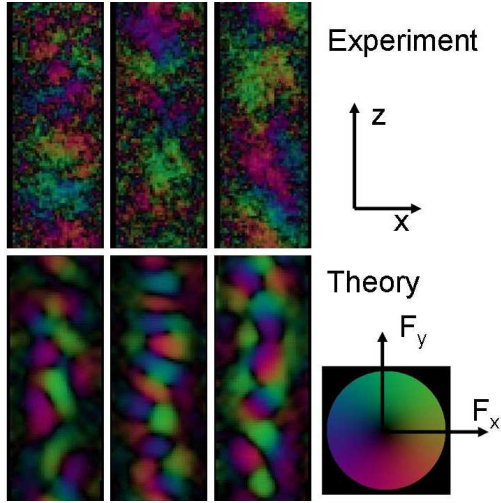}
\caption{Spatial structure of the complex transverse magnetization
$F_\perp(\mathbf{r})$ in the $x-z$ plane at $t = 87$ ms for $q/h = 2$ Hz.
 The upper panel of figures consist of random instances of experimentally measured spin textures\cite{dmsk_prl} while the lower figures are random instances of calculated spin textures.
 Each figure in the panel is 100 $\mu$m long and 40 $\mu$m wide.
The complex number $F_\perp$ is represented as shown in the color wheel in the inset of the figure. Domains are seen to have limited
 structure
along the width of the trap and localized
at the center of the condensate. }
\end{figure}

\subsection{Effect of  the dipole interaction}

As discussed in  Section VI,  dipole-dipole interactions
reduce the rate of domain formation in the case where the magnetic field is aligned along the $z$-direction, which is the long
axis of the condensate. However, as seen in Fig.\ 5, for the parameters of the calculation, which are taken to be the ones
relevant to experiment, the effect of the dipole-dipole
interaction on the average transverse magnetization $G(\mathbf{0})$ turns out to be
small because the Fourier transform of the dipole interaction kernel $K$ almost vanishes at the
lengthscale of domain formation. That is, the spin healing length being nearly equal
to the narrower condensate thickness, the dominant length scale for domain formation coincides with the cross-over between the 2D and 3D forms of the dipole
interaction. Despite having a negligible effect on domain formation in the longitudinal direction,
 dipole-dipole interactions are found to suppress
domain formation along the radial direction.

\begin{figure}\label{fig:G0dip_dir}
\centering
\includegraphics[scale=0.4]{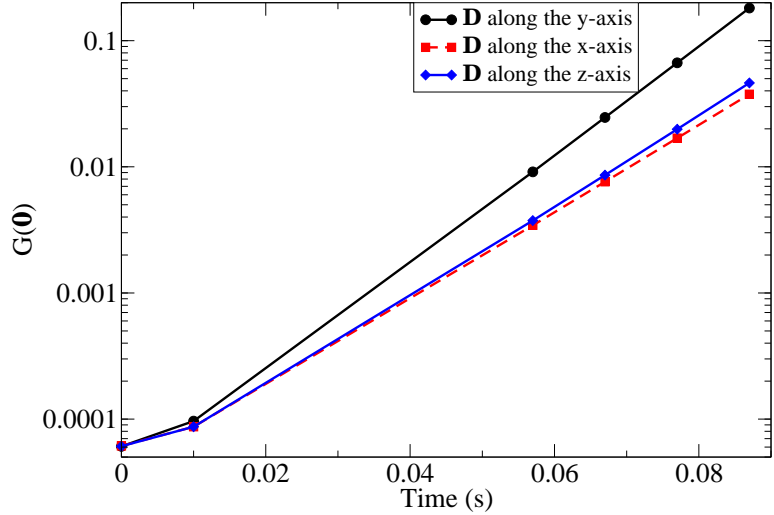}
\caption{Dependence of the evolution of the variance of the transverse magnetization for the dipole-precession axis $\mathbf{D}$
 aligned along $\hat{\mathbf{x}}$,$\hat{\mathbf{y}}$ and $\hat{\mathbf{z}}$ directions for $q/h=2.0$ Hz. The growth rate of transverse magnetization is
found to be enhanced significantly for $\mathbf{D}$ aligned along $\hat{\mathbf{y}}$ as compared to $\mathbf{D}$ aligned along $\hat{\mathbf{z}}$
which yields results close to the case without dipole interaction.}
\end{figure}

Yet as discussed in Section VI, other experimental geometries, i.e.\ orientations of $\mathbf{D}$ and $\mathbf{P}$ away from the
$z$ axis, are expected to show more prominent dipolar effects in the spontaneous formation of magnetization.
We explored this possibility numerically. The magnitude of the magnetization variance
$G(\mathbf{0})$  indicated by such calculations
is shown in Fig.\ 6.
One can see that the rate of growth of transverse magnetization  is significantly enhanced with $\mathbf{D}$ and $\mathbf{P}$
pointing along the  $y$-direction compared to other orientations.

\section{Conclusion}

 In our analysis we have studied a realistic quantum Hamiltonian model for a quasi-two dimensional
spinor condensate, including the effects of the trap and dipole-dipole interactions, in order to make a
quantitatively accurate prediction of the contribution of intrinsic fluctuations to the symmetry-breaking
domain formation.
   Similar to previous studies, the inclusion of the trapping potential was found to reduce the rate of structure
 formation
 because of a reduction of the average density. The dipole-dipole interaction, which is known to have a dominant effect
on the long term structure formation in spinor BECs, was found to add an effective
non-local contribution to the spin-dependent part of the interaction in the spinor condensate. The non-local
nature of the spin-spin interactions couples the spin structure formation dynamics to the direction
of the spin polarization.
Even though dipole-dipole interactions are found to affect the spin dynamics weakly when $\mathbf{D}$ and $\mathbf{P}$
are polarized along $m_z=0$, we find that dipole interactions significantly enhance the rate of domain formation when these
vectors are  polarized along $m_y=0$. Moreover, dipole-dipole interactions
were found to split the degeneracy of the two polarizations of the magnon modes
in the case where $\mathbf{D}$ was orthogonal to $\mathbf{P}$. This
spin-polarization dependence of the domain formation rate leads to a direct way to
observe experimentally
the role of dipole-dipole interaction on spinor dynamics.

\begin{figure}
\centering
\includegraphics[scale=0.4]{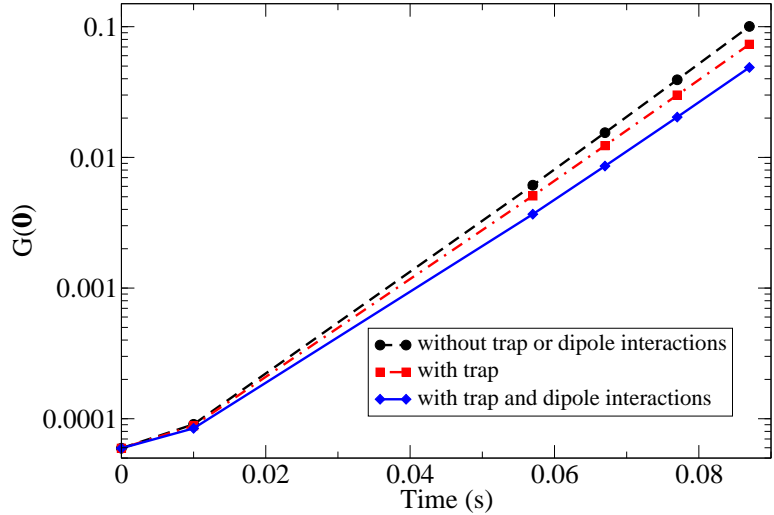}
\caption{Summary of effects of various factors on the evolution of the magnitude of transverse magnetization.Both dipole-dipole interactions and
the external trapping interaction are found to reduce the growth rate of the transverse magnetization.}
\end{figure}

 A quantitative prediction of the magnitude of the structure formation also requires the inclusion
of effects from non-linear interaction terms and thermal effects. On a preliminary examination one would have expected
thermal effects to be significant since the kinetic temperature of the condensate is much larger than the spin-mixing
energy scale. However we found that the coupling of phonons to spin fluctuations is small,
 leading to a separation of the low temperature spin dynamics from the high temperature phonon dynamics. In Section IV, we studied the effects of the
 non-linear interactions within the standard TWA and found
that non-linearity effects lead to saturation of the transverse magnetization at long times. We expect the TWA
to be a reasonably accurate description of the spin-spin correlations of a spinor BEC since it was found to yield
 results in good agreement with exact diagonalization calculations our spinor BEC model in the single mode
regime.

Despite our effort to include the effects of the trapping potential, dipole-dipole interactions, non-linearities and finite
temperature to develop a quantitative understanding of the magnitude of domain formation, we found in Section VII that
the uncertainty in the magnitude of the spin-dependent part of the contact interaction prevents us from making a quantitative
comparison of the magnitude of the domain formation with experimental results. Such a quantitative comparison between theory
and experiment is critical for the determination of the contribution of intrinsic quantum fluctuations to the domain
formation. One possible experimental approach  to resolving
this problem is to determine in a direct way the gain of the spinor BEC in the
experimental geometry by studying the dynamics of the magnetization of the
condensate following an initial microwave pulse. Such experiments
in conjunction with quantitative calculations might make it possible
 to determine better the importance of intrinsic quantum fluctuations to symmetry breaking dynamics.

This work was supported by the NSF, the U.S Department of Energy under Contract No. DE-AC02-05CH11231, DARPA's OLE Program,
 and the LDRD Program at LBNL. S. R. L.  acknowledges support from the NSERC.
Computational resources have
been provided by NSF through TeraGrid resources at SDSC,
DOE at the NERSC, TACC, Indiana University.

\appendix
\section{ Eigenmodes for the Bogoliubov transformation of inhomogeneous dipolar condensates.}

Here we give explicit expressions for the eigenmodes
 and eigenfrequencies of a dipolar ferromagnetic spinor BEC
for positive quadratic Zeeman shifts. As discussed in Section VI, the inclusion of dipole-dipole interactions requires
the generalization of the local spin-dependent coupling constant $c_2 n$ to $c_2^{\mbox{\scriptsize{eff}}}(\mathbf{r_1},\mathbf{r_2})$. The eignmodes
and eigenfrequencies that we  define are strictly  valid when the Hermitean Hamiltonian
 $H_{0}=-\frac{\hbar^2}{2 M}\nabla^2+q(t)+\mu+c_0 n(\mathbf{r})+V_{trap}$ is positive definite.
In this case the eigenmodes of the condensate are given by
\begin{align}
\Upsilon_{+1}^{(\sigma n)}(\mathbf{r})&=\sigma E_n^{-1}H_0^{1/2}\xi_n(\mathbf{r})\\
\Upsilon_{-1}^{(\sigma n)}(\mathbf{r})&=H_0^{-1/2}\xi_n(\mathbf{r}).
\end{align}
where  $E_n^2$ and $\xi_n$ are defined to be eigenvectors and eigenvalues of the
 Hermitean operator $H_0^{1/2}(H_0+2 c_2^{\mbox{\scriptsize{eff}}})H_0^{1/2}$ and $\sigma=\pm 1$.
The dual modes then follow to have the form
\begin{align}
\tilde{\Upsilon}_{+1}^{(\sigma n)}(\mathbf{r})&=\sigma E_n H_0^{-1/2}\xi_n(\mathbf{r})\\
\tilde{\Upsilon}_{-1}^{(\sigma n)}(\mathbf{r})&=H_0^{1/2}\xi_n(\mathbf{r}).
\end{align}

In the case of negative quadratic Zeeman shifts,  such a Hermitean eigenproblem cannot be constructed
since some of the frequencies in this case are neither purely real nor imaginary. This can be verified by
introducing a weak periodic potential to the homogeneous ferromagnetic Bose gas
and diagonalizing the problem using degenerate perturbation theory.

\section{Regularizing the dipole potential}

In order to perform numerical calculations, even semi-analytic calculations where the Fourier transform
of the three-dimensional dipole interaction kernel, $K(\mathbf{R}-\mathbf{R}')=\frac{(\mathbf{R}-\mathbf{R}')^2-3(\mathbf{D}\cdot(\mathbf{R}-\mathbf{R}'))^2}{|\mathbf{R}-\mathbf{R}'|^5}$, is needed, one needs the integral involved in the
 Fourier transform of
the kernel to be well defined.
 The full 3D Fourier transform of the dipole interaction kernel may be calculated
 analytically,
 but to obtain converged
results for the spin dynamics it is necessary to truncate the long-ranged dipole interaction
 between periodic images
of the system which emerge when using Fourier techniques to do such calculations.
 For numerical convenience
we imagine that the dipole density can be expanded in terms of a possibly overcomplete set of functions i.e
\begin{equation}
\phi(\mathbf{R})=\frac{\sum_n \phi(\mathbf{R}^{(grid)}_n) \rho(\mathbf{R}-\mathbf{R}^{(grid)}_n)}{\sum_{n}\rho(\mathbf{R}^{(grid)}_n)}.
\end{equation}
Such an expansion allows us to represent a function $\phi(\mathbf{R})$ which is smooth on the scale of the width
of $\rho(\mathbf{R})$ by its value $\phi(\mathbf{R}^{(grid)}_n)$ on a discrete grid of points $\mathbf{R}^{(grid)}_n$ with a
 grid spacing
 that is smaller than
the width of $\rho(\mathbf{R})$.
Integrals of the kernel $K(\mathbf{R})$ of interest are given by
\begin{align}
\int
\phi(\mathbf{R}_1)\phi(\mathbf{R}_1)K(\mathbf{R}_1-\mathbf{R}_2)&=\sum_{n_1,n_2}\frac{\phi(\mathbf{R}^{(grid)}_{n_1})\phi(\mathbf{R}^{(grid)}_{n_2})}{(\sum_{n}\rho(\mathbf{R}^{(grid)}_n))^2}\\
&\int
\,\ud\mathbf{R}'_1
\,\ud\mathbf{R}'_2\rho(\mathbf{R}'_1-\mathbf{R}_1)\rho(\mathbf{R}'_2-\mathbf{R}_2)K(\mathbf{R}_1-\mathbf{R}_2). \end{align}

Taking the smoothing function to be $\rho(\mathbf{R})=\frac{1}{w^3(2\pi)^{(3/2)}}e^{-R^2/2 w^2}$,
the averaged kernel $g$ is given by
\begin{align}
g(\mathbf{R})=\int d\mathbf{R}_1 d\mathbf{R}_2 \,\rho(\mathbf{R}_1-\mathbf{R})\rho(\mathbf{R}_2)K(\mathbf{R}_1-\mathbf{R}_2)=
\frac{(3 \cos^2(\theta)-1)}{R^3}\\
&\left[\textrm{Erf}\left(\frac{R}{2 w}\right)-\frac{R}{w\sqrt{\pi}}e^{-\frac{R^2}{4 w^2}}\left(\frac{R^2}{6 w^2}+1\right)\right].
\end{align}
 The regularized expression, unlike the original dipole interaction kernel $K(\mathbf{R})$, vanishes for small $R$ and approaches the regular
 expression $\frac{(3 \cos^2(\theta)-1)}{R^3}$ for
large $R$ as expected.

Using the regularized kernel $g(\mathbf{R})$, the dipole interaction can  be calculated on a real-space grid with grid spacings smaller than the width of the smoothing profile
 $\rho(\mathbf{R})$,
 in a way so as to avoid interaction
between periodic images. The Fourier transform for the two dimensional kernel with a Gaussian profile given Eq.\ \ref{eq:F} in Section VI
can be derived by applying a Fourier transform to $g(\mathbf{R})$ restricted to the 2D plane.

\bibliography{reference}
\newpage

\end{document}